\documentclass[12pt]{article}

\usepackage{amsmath}
\usepackage{newtxtext}
\usepackage[subscriptcorrection]{newtxmath}
\usepackage{bm}
\usepackage{comment}
\usepackage{graphicx}
\usepackage[plain,noend]{algorithm2e}
\usepackage{apacite}  
\usepackage{geometry} 
\usepackage{natbib}
\newtheorem{theorem}{Theorem}
\usepackage{authblk}

\makeatletter
\renewcommand{\algocf@captiontext}[2]{#1\algocf@typo. \AlCapFnt{}#2} 
\def\@algocf@capt@plain{top}
\renewcommand{\algocf@makecaption}[2]{%
  \addtolength{\hsize}{\algomargin}%
  \sbox\@tempboxa{\algocf@captiontext{#1}{#2}}%
  \ifdim\wd\@tempboxa >\hsize
    \hskip .5\algomargin%
    \parbox[t]{\hsize}{\algocf@captiontext{#1}{#2}}
  \else%
    \global\@minipagefalse%
    \hbox to\hsize{\box\@tempboxa}
  \fi%
  \addtolength{\hsize}{-\algomargin}%
}
\makeatother

\def\T{{ \mathrm{\scriptscriptstyle T} }}

\begin{document}

\title{\bf A measure of departure from symmetry via the Fisher-Rao distance for contingency tables}

\author[1]{Wataru Urasaki}
\author[1]{Go Kawamitsu}
\author[2]{Tomoyuki Nakagawa}
\author[1]{Kouji Tahata}

\affil[1]{Department of Information Sciences, Tokyo University of Science}
\affil[2]{School of Data Science, Meisei University}
\date{}

\maketitle

\begin{abstract}
A measure of asymmetry is a quantification method that allows for the comparison of categorical evaluations before and after treatment effects or among different target populations, irrespective of sample size. 
We focus on square contingency tables that summarize survey results between two time points or cohorts, represented by the same categorical variables. 
We propose a measure to evaluate the degree of departure from a symmetry model using cosine similarity. 
This proposal is based on the Fisher-Rao distance, allowing asymmetry to be interpreted as a geodesic distance between two distributions.
Various measures of asymmetry have been proposed, but visualizing the relationship of these quantification methods on a two-dimensional plane demonstrates that the proposed measure provides the geometrically simplest and most natural quantification.
Moreover, the visualized figure indicates that the proposed method for measuring departures from symmetry is less affected by very few cells with extreme asymmetry. 
A simulation study shows that for square contingency tables with an underlying asymmetry model, our method can directly extract and quantify only the asymmetric structure of the model, and can more sensitively detect departures from symmetry than divergence-type measures.
\end{abstract}

\medskip

{\bf Keywords}: Contingency table, Symmetry model, Measure of asymmetry, Cosine similarity, Fisher-Rao distance

\medskip

{\bf Mathematics Subject Classification}: 62H17, 62H20

\section{Introduction}
When evaluating treatment effects in repeated measures or assessing the diagnostic accuracy of diseases using categorical variables, it is effective to analyze square contingency tables that cross-tabulate these evaluations.
Two-way square contingency tables, comprising two categorical variables, have led to numerous proposals by focusing on the ``symmetry'' structure, which reflects category transitions. 
Although a classical method was proposed over half a century ago, McNemar's test, introduced by \cite{mcnemar1947note}, remains applicable today. 
The McNemar's test is highly useful for comparing treatments evaluated with binary categories, and its simplicity and broad applicability have made it widely used in fields such as healthcare, education, and marketing.
For multi-valued category cases, \cite{bowker1948test} proposed the Bowker's test by introducing a symmetry (S) model, which extends the McNemar's test, and it is also widely utilized.
Additionally, the Cochran-Mantel-Hansel's test, proposed by \cite{mantel1959statistical}, aims to achieve precise effect measurement by stratifying categorical confounding factors that influence evaluations, and it continues to make significant contributions.

While the introduced methods are innovative, they remain statistical tests limited to determining the presence or absence of the treatment effect or transitions at a significance level of $\alpha\%$. 
Therefore, there is a method known as the measure of asymmetry for quantifying the degree of the effect or transitions within a fixed interval, regardless of sample size.
The measure is quantified by defining the maximum deviation structure from the S model. 
Consequently, dividing the table by confounding factors and analyzing each with the measure enables comparisons across confounding factors, which statistical tests could not achieve. 
Various proposals have been made; for instance, \cite{tomizawa1994two} proposed measures based on Kullback-Leibler divergence and Pearson divergence, and \cite{tomizawa1998power} introduced a generalization of divergence-type measures using power divergence. 
However, since divergence itself is a pseudo-distance, a measure based on Matusita distance (also known as Hellinger distance) was proposed by \cite{yamamoto2008distance}, following the work of \cite{matusita1953estimation, matusita1955decision}. 
Additionally, \cite{tahata2009measure} proposed a measure based on the angle that reflects the degree of asymmetry.
While numerous quantification methods have been proposed, determining the optimal measure remains a subject of ongoing debate.

We propose a measure of asymmetry based on cosine similarity for two-way square contingency tables with nominal categories, under the Fisher–Rao distance to measure the degree of departure from symmetry.
The Fisher-Rao distance provides the difference between probability distributions in closed form, which eliminates the need for complex numerical analysis and allows for direct calculation of the distance, offering practical advantages for applications.
This approach also relies on a square-root representation, which enables the analytical computation of geodesic paths and distances, allowing the degree of asymmetry to be interpreted as the geodesic distance in terms of the arc length on the unit circle.
The primary objective of this paper is to present a measure that provides the simplest and most intuitive quantification for the departure from symmetry, while minimizing the impact of extreme asymmetry with a small number of cell probabilities.
Additionally, we visualize various quantification methods represented by coordinates in the first quadrant where both the x- and y-axes range from $0$ to $1$, to demonstrate their comparative performance.
From the visualization, our proposal is shown as a natural quantification method statistically based on the geometric relationship and information geometry.
Moreover, a simulation study demonstrates the ability to extract and quantify only the asymmetric structure and to sensitively detect and compare slight departure from symmetry than divergence-type measures.
In the Appendix , we introduce examples of real data analysis and proofs of several properties related to the proposed measure.

\section{Notation}
\subsection{S model for square contingency table}
Consider an $R \times R$ square contingency table that has the same row and column classification with nominal categories.
Let $p_{ij}$ denote the probability that an observation will fall in the table's $i$th row and $j$th column ($i = 1, \dots,R; j = 1,\dots,R$).
\cite{bowker1948test} considered the S model defined as
\begin{equation*}
p_{ij} = p_{ji}; \; i, j = 1, \dots, R.
\end{equation*}
Additionally, it can also be expressed by the conditional probability $p^c_{ij} = p_{ij}/(p_{ij} + p_{ji})$:
\begin{equation*}
p^c_{ij} = \frac{1}{2}; \; i, j = 1, \dots, R.
\end{equation*}
The S model assumes that the probability that an observation will fall in cell $(i,j)$ is equal to the probability that it falls in symmetric cell $(j, i)$ for $i\neq j$.
Under this hypothesis, Bowker's $\chi^2$ statistics $\chi^2_{S}$ follows a chi-squared distribution with $R(R-1)/2$ degrees of freedom.
Bowker's test is a generalization of McNemar's test (\citealp{mcnemar1947note}) for $R\times R$ contingency table with $R>2$.
To quantitatively evaluate the degree of departure from symmetry when the S is rejected by the test, measures of asymmetry have been proposed using various quantification methods.

\subsection{Quantification methods}
As approaches for detecting differences, numerous quantification methods, such as distances, similarities, and pseudo-distances, have been proposed across many fields to date.
Let ${p_i}$ and ${q_i}$ $(i = 1, \dots, l)$ be probability vectors from two discrete probability distributions. 
Among the many methods, the four that are most relevant to this study are as follows.
(I) Euclidean distance: $Ed(\{p_i\}, \{q_i\})$, (II) Hellinger distance: $Hd(\{p_i\}, \{q_i\})$, (III) Fisher-Rao distance: $FRd(\{p_i\}, \{q_i\}) = \cos^{-1}\left[Sc(\{\sqrt{p_i}\}, \{\sqrt{q_i}\})\right]$, and (IV) Power-divergence: $Pd^{(\lambda)}(\{p_i\};\{q_i\}) = \{\lambda(\lambda-1)\}^{-1}\sum^{l}_{i=1} p_i\{(p_i/q_i)^{\lambda} - 1\}$ for $-\infty < \lambda < \infty$.
Note that $Sc(\{\sqrt{p_i}\}, \{\sqrt{q_i}\})$ represents the cosine similarity.
$Ed(\{p_i\}, \{q_i\})$ is a straightforward measurement method as it measures the direct distance between the vectors. 
$Hd(\{p_i\}, \{q_i\})$, on the other hand, is used when there are outliers present, minimizing their impact on the measurement.
$Sc(\{\sqrt{p_i}\}, \{\sqrt{q_i}\})$ depends solely on the angle $\theta$ between the two square-root transformed vectors of $\{p_i\}$ and $\{q_i\}$, making it useful for measuring similarity.
It is primarily used in machine learning to measure the similarity between words within text data. 
The cosine similarity  takes values ranging from $-1$ to $1$, where the value of $1$ indicates that the two vectors are similar, and the value of $0$ indicates no relation.
Conversely, the value of $-1$ means that they are completely dissimilar.
Thus, $FRd(\{p_i\}, \{q_i\})$ represents the arc length on the unit circle corresponding to the angle formed by the two square-root transformed vectors.
Furthermore, this method corresponds to a Fisher-Rao distance for discrete probability distributions and possesses several statistically advantageous properties for detecting differences. 
For more details on the Fisher-Rao distance and its advantages, please refer to \cite{miyamoto2024closed}, \cite{calin2014geometric}, \cite{atkinson1981rao}, and \cite{burbea1984informative}.
These three quantification methods satisfy the axioms of distance and directly indicate the difference between the vectors.
On the other hand, $Pd^{(\lambda)}(\{p_i\};\{q_i\})$ does not satisfy the axioms of distance but is used to measure the difference between distributions.
For further information on divergence, please refer to \citealp{renyi1961measures}, \citealp{read2012goodness}, and \citealp{cressie1984multinomial}.
The measurement method varies depending on the parameter provided.
When $\lambda = -1/2$, $0$, $1$, it can be expressed as follows: (IV-a) Hellinger distance: $Pd^{(-1/2)}(\{p_i\};\{q_i\})$, (IV-b) Kullback-Leibler divergence: $Pd^{(0)}(\{p_i\};\{q_i\})$, and (IV-c) Pearson divergence: $Pd^{(1)}(\{p_i\};\{q_i\})$.
It should be noted that both $Hd(\{p_i\}, \{q_i\})$ and $Pd^{(-1/2)}(\{p_i\};\{q_i\})$ are referred to as ``Hellinger distance'', however, the former measures the distance between probability vectors while the latter measures the difference between distributions.
Thus, $Hd(\{p_i\}, \{q_i\}) \neq Pd^{(-1/2)}(\{p_i\};\{q_i\})$, and distinguishing their notations and use.

\section{Measure of asymmetry based on the Fisher-Rao distance}
\subsection{Proposed measure via the Fisher-Rao distance}
When considering the conditional probability $p_{ij}^{c}$,
we then obtain a restriction $p_{ij}^{c}+p_{ji}^{c}=1$, and applying the square-root transformation leads to the condition $\{(p_{ij}^{c})^{1/2}\}^2+\{(p_{ji}^{c})^{1/2}\}^2=1$.
A vector $\bm{p}_c=(p_{ij}^{c}, p_{jj}^{c})^\T$ can then be interpreted as a coordinate representing the degree of asymmetry under this restriction.
Additionally, when the vector $\bm{p}_c$ coincide with $\bm{s}=(1/2, 1/2)^\T$, it indicates that $p_{ij}^c = p_{ji}^c= 1/2$ (i.e., these cells $(i,j)$ and $(j,i)$ have symmetry).

Assuming that $\{p_{ij} + p_{ji} \neq 0 \}$, we propose a cosine similarity type measure to represent the degree of departure from symmetry as follows:
\begin{equation*}
\Phi = \frac{4}{\pi} \mathop{\sum \sum}_{i \neq j} w_{ij}(\bm{p}) FRd \left( \bm{p}_c , \bm{s} \right),
\end{equation*}
where $\bm{p} = (p_{11},\dots,p_{1R},\dots,p_{R1},\dots,p_{RR})^\T$, and $w_{ij}(\bm{p})$ is the arbitrary weight satisfied with
\begin{equation*}
w_{ij}(\bm{p}) > 0, \quad \mathop{\sum \sum}_{i \neq j} w_{ij} (\bm{p} ) = 1 .
\end{equation*}
The proposed measure $\Phi$ can also be simplified as
\begin{equation}\label{eq:measure}
\Phi = \frac{4}{\pi} \mathop{\sum \sum}_{i \neq j} w_{ij}(\bm{p}) \cos^{-1} \left( \frac{\sqrt{p_{ij}} + \sqrt{p_{ji}}}{\sqrt{2( p_{ij} + p_{ji})}} \right).
\end{equation}
The weight $w_{ij}(\bm{p})$ should be chosen by the user's analysis objective, some examples are $1/\{R(R-1)\}$, $(p_{ij}+p_{ji})/(2\sum_{i\neq j} p_{ij})$, and so on.
In particular, when using the weight $1/\{R(R-1)\}$, the measure's value represents the arithmetic mean of the degree of departure from symmetry for each cell. 
Therefore, if simple and intuitive results are desired, this weight is recommended.
Additionally, the proposed measure $\Phi$ has the following three properties:
\begin{theorem}\label{the1}
The measure $\Phi$ satisfies the following three properties:
\begin{itemize}
\item[1] $\Phi$ must lie between 0 and 1.
\item[2] $\Phi = 0$ if and only if there is a complete structure of symmetry, i.e., $p_{ij}=p_{ji}$.
\item[3] $\Phi = 1$ if and only if there is a structure in which the degree of departure of symmetry is the largest, i.e., $p_{ij} = 0$ (then $p_{ji} > 0$) or $p_{ji} = 0$ (then $p_{ij} > 0$).
\end{itemize}
\end{theorem}
The proof of Theorem \ref{the1} is given in the Appendix .

\subsection{Geometric relationships for quantification methods}
Among the various quantification methods, our proposal utilizes cosine similarity to suggest a new measure. 
In quantifying asymmetry, we demonstrate that a method based on cosine similarity is the most effective by examining its geometric relationships with other methods.

\begin{figure}[htbp]
\centering
\includegraphics[width=0.8\linewidth]{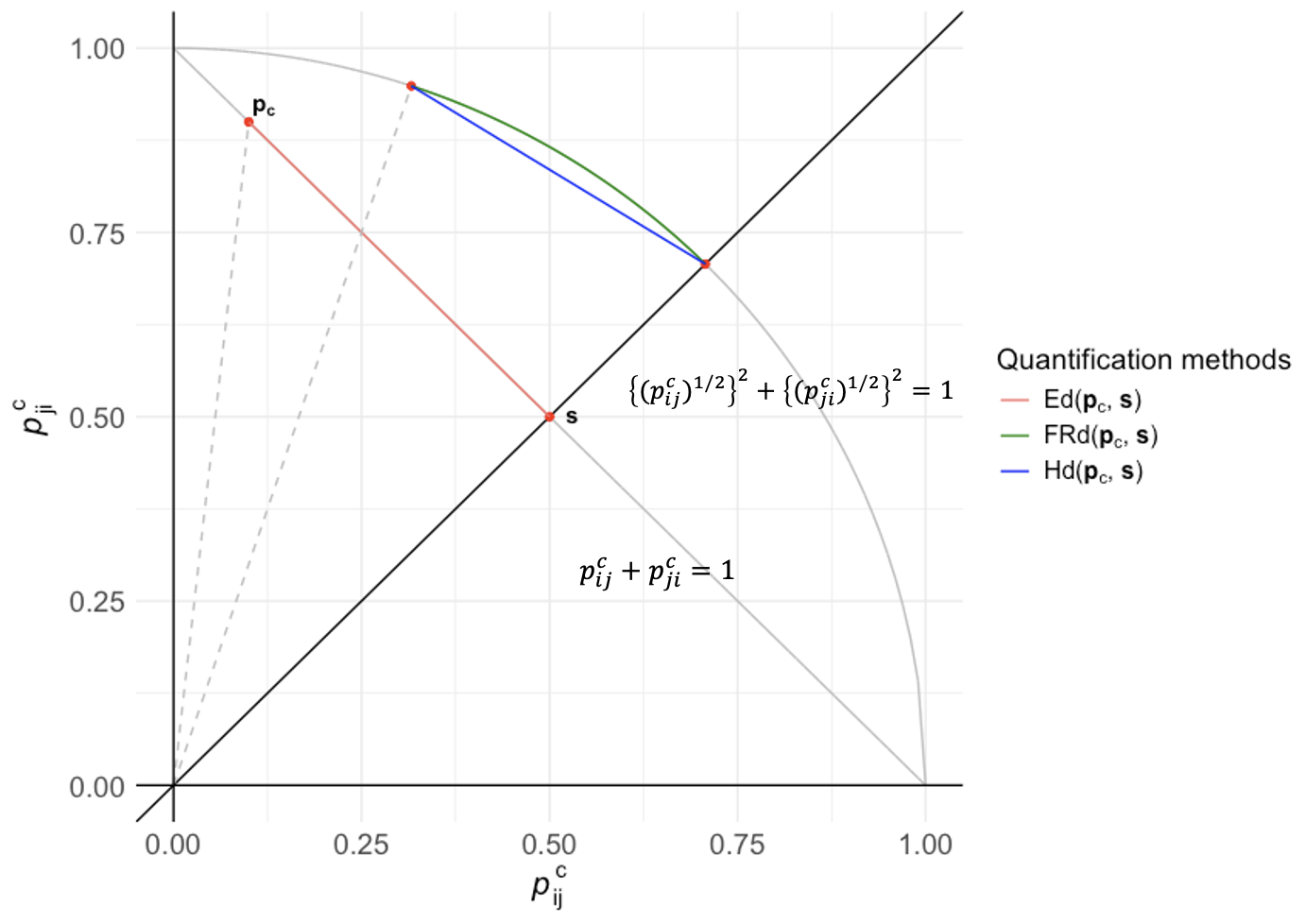}
\caption{Example for geometric relationships between three quantification methods.}
\label{plot1}
\end{figure}

For any cell $(i,j)$ in a square contingency table, the departure from symmetry can be represented by the vector $\bm{p}_c$.
When perfect symmetry is present, the vector coincides with $\bm{s}$.
To quantify the asymmetry of cell $(i,j)$, we can get three important distances $Ed(\bm{p}_c, \bm{s})$, $FRd(\bm{p}_c, \bm{s})$, and $Hd(\bm{p}_c, \bm{s})$.
Certainly, other types of distances could also be considered. 
However, they tend to be computationally difficult or their interpretation as a distance is complex, making them unsuitable for practical applications.
Therefore, we have focused on these distances due to their simplicity and intuitive nature as quantification methods.
The methods are illustrated in Fig.\ref{plot1}.

The Fig.\ref{plot1} plots $p^c_{ij}$ on the x-axis and $p^c_{ji}$ on the y-axis, illustrating the distances $Ed(\bm{p}_c, \bm{s})$, $FRd(\bm{p}_c, \bm{s})$, and $Hd(\bm{p}_c, \bm{s})$.
Examining this figure, it becomes apparent that there could be many ways to quantify the distance between vectors other than those discussed in this paper. 
However, under the restriction $p_{ij}^{c}+p_{ji}^{c}=1$ and the square-root transformation $\{(p_{ij}^{c})^{1/2}\}^2+\{(p_{ji}^{c})^{1/2}\}^2=1$, only feasible $Ed(\bm{p}_c, \bm{s})$ and $FRd(\bm{p}_c, \bm{s})$ are easy to calculate and intuitive.
Furthermore, $Hd(\bm{p}_c, \bm{s})$ is also considered an easily interpretable and intuitive quantification method. 
However, as can be visually observed from Fig.\ref{plot1}, it cannot be said to measure accurately when taking the constraints into account.
Next, we examine the relationship between $Ed(\bm{p}_c, \bm{s})$ and $FRd(\bm{p}_c, \bm{s})$ in terms of their magnitude.
Although determining the exact magnitude relationship between $Ed(\bm{p}_c, \bm{s})$ and $FRd(\bm{p}_c, \bm{s})$ is theoretically challenging, computational results indicate that within the range $0.005 < p^c_{ij} < 0.995$, $FRd(\bm{p}_c, \bm{s}) < Ed(\bm{p}_c, \bm{s})$.
It can be said that, except in cases that nearly exhibit asymmetry ($p^c_{ij} = 0$ or $1$), $FRd(\bm{p}_c, \bm{s})$ will generally be smaller.
This finding suggests that the proposed measure $\Phi$, which is derived from $FRd(\bm{p}_c, \bm{s})$, provides an assessment that reduces the influence of a small number of cell probabilities exhibiting extreme asymmetry.

While we have discussed the usefulness of adopting the square-root transformation and $FRd(\bm{p}_c, \bm{s})$ in comparison to various distances, there are additional benefits to the two as described by \cite{kurtek2015bayesian}.
The notable benefit is that the transformation allows the degree of asymmetry to be treated as a geodesic distance under the Fisher-Rao distance for the Bernoulli distribution in $p^c_{ij}$.
Since the geodesic distance can be easily defined as an angle, the degree of asymmetry can be obtained by a closed-form solution without relying on complex numerical methods. 
Therefore, the simplicity of the calculation provided by the transformation proves to be highly useful for applications and is especially advantageous in the case of multi-way square contingency tables.
Additionally, the fact that $FRd(\bm{p}_c, \bm{s})$ represents the Fisher-Rao distance ensures consistency in distance evaluation, even when different parameterizations of the two vectors are used.

We next consider the comparison between arc length and power-divergence.
Since these quantify different aspects (distance metrics versus divergence), a direct comparison is not straightforward. 
Therefore, in the next section, we investigated the differences in performance and characteristics as the measure of asymmetry by comparing our proposed measure $\Phi$, based on $FRd(\bm{p}_c, \bm{s})$, with $\Phi^{(\lambda)}$, which was proposed by \cite{tomizawa1998power} and is based on the power-divergence.
However, from the outset, the proposed measure $\Phi$ can be considered as an average distance for the asymmetry of each cell in the contingency table, indicating that it has greater interpretability than the power-divergence-type measure $\Phi^{(\lambda)}$.

\section{Simulation experiment}
To investigate the differences in performance and characteristics between our measure $\Phi$ and the power-divergence-type measure $\Phi^{(\lambda)}$ in quantifying asymmetry, we consider a probability table with the structure of the conditional symmetry (CS) model by \cite{mccullagh1978class}, which is one of the asymmetry models. 
The CS model is defined as
\begin{equation*}
p_{ij} = \left\{
\begin{array}{rl}
\Delta \psi_{ij} & (i < j),\\
\psi_{ij} & (i \geq j),
\end{array}
\right.
\end{equation*}
where $\Delta > 0$ and $\psi_{ij} = \psi_{ji}$. 
Additionally, $\delta \rightarrow 0$ is treated as equivalent to $\delta = 0$.
According to the model, $\Delta = 0$ or $1$ signifies that the probability table represents a perfectly asymmetric or symmetric structure, respectively.
Conversely, when $\Delta > 1$, the cell probabilities in the upper triangular part increase, while for $\Delta < 1$, the opposite pattern is observed.
By systematically varying the degree of departure from symmetry within the range $0 \leq \Delta \leq 1$, we explore the characteristics of the two types of measures.
For the power-divergence-based measure, we consider parameter values of $\lambda = -1/2, 0$, and $1$.

\begin{figure}[htbp]
\centering
\includegraphics[width=0.8\linewidth]{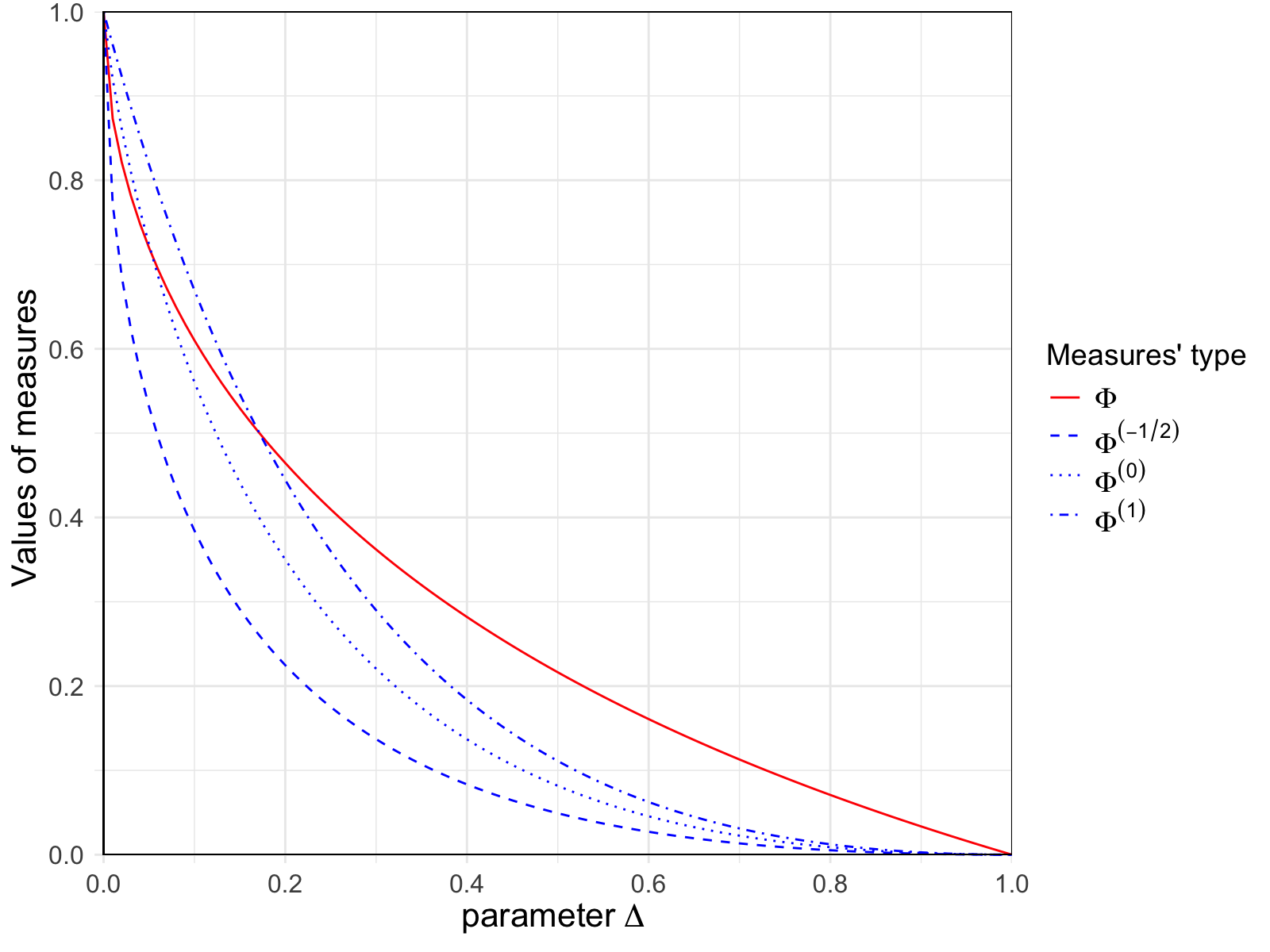}
\caption{$\Phi$ versus $\Phi^{(\lambda)}$ with $\lambda=-1/2, 0$, and $1$ under $0 \leq \Delta \leq 1$}
\label{plot3}
\end{figure}

\begin{table}[htp]
\centering
\caption{Values of $p^c_{ij}$, $(p^c_{ij})^{1/2}$, $\Phi$ and $\Phi^{(\lambda)}$ for each $\Delta$ }
{\begin{tabular}{crrcrrrr}
& \multicolumn{2}{c}{Coordinates} && \multicolumn{4}{c}{Two type measures}  \\
$\Delta$ & \multicolumn{1}{c}{$p^c_{ij}$} & \multicolumn{1}{c}{$(p^c_{ij})^{1/2}$} & & \multicolumn{1}{c}{$\Phi$} & \multicolumn{1}{c}{$\Phi^{(-1/2)}$} & \multicolumn{1}{c}{$\Phi^{(0)}$} & \multicolumn{1}{c}{$\Phi^{(1)}$} \\ 
$0$ & $1$ & $1$ & & $1$ & $1$ & $1$ & $1$ \\
$0.2$ & $0.883$ & $0.913$ & & $0.465$ & $0.225$ & $0.350$ & $0.444$ \\
$0.4$ & $0.714$ & $0.845$ & & $0.282$ & $0.083$ & $0.137$ & $0.184$ \\
$0.6$ & $0.625$ & $0.791$ & & $0.161$ & $0.027$ & $0.046$ & $0.062$ \\
$0.8$ & $0.556$ & $0.745$ & & $0.071$ & $0.005$ & $0.009$ & $0.012$ \\
$1$ & $0.5$ & $0.707$ & & $0$ & $0$ & $0$ & $0$ \\ 
\end{tabular}}
\label{tablelabel1}
\end{table}

Figure \ref{plot3} displays $\Phi$ versus $\Phi^{(\lambda)}$ for $0 \leq \Delta \leq 1$, and Table \ref{tablelabel1} shows the partial results of Fig.\ref{plot3} along with the corresponding $p^c_{ij}$ and $(p^c_{ij})^{1/2}$.
It is important to note that the values of $\Phi$ are unaffected by weights.
When the CS model can be assumed completely, (\ref{eq:measure}) can be expressed as
\begin{equation*}
\Phi = \frac{4}{\pi} \cos^{-1} \left( \frac{1 + \sqrt{\Delta}}{\sqrt{2(1 + \Delta)}} \right).
\end{equation*}
The equation shows that the departure from symmetry depends solely on the asymmetric structure $\Delta$, irrespective of weights.
This is because $\Phi$  can be expressed only by an odds $\{p_{ij}/p_{ji}\}$ or $\{p_{ji}/p_{ij}\}$, which isolates the asymmetric structure.
If in the analysis of real data, the values of $\Phi$ do not change with varying weights, it suggests the presence of a simple asymmetry model structure, including the CS model.

Next, let’s examine the differences in the values' performance of the two type measures using both Fig.\ref{plot3} and Table \ref{tablelabel1}. 
The proposed measure $\Phi$ shows higher values than the existing measure $\Phi^{(\lambda)}$ even when $\Delta$ is near $1$ (indicating small departures from symmetry), allowing it to detect subtle asymmetry structures that might be missed. 
This improves the ability to identify symmetrical structures in the data and captures slight asymmetries more effectively than divergence.
Conversely, when $\Delta$ is close to $0$ (indicating large asymmetry), the value of $\Phi$ is lower than $\Phi^{(\lambda)}$ for some parameters, preventing overestimation of extreme asymmetry.
This allows for more appropriate evaluation of the analysis results and reduces the risk of overestimation.
Furthermore, the proposed measure shows a gradual consistent decrease over the entire range of $\Delta$, allowing for stable measurement of asymmetry regardless of the intensity of the asymmetry. 
While divergence measures rapidly increase as the departure from symmetry becomes stronger, our proposal does not exhibit such abrupt changes, ensuring consistency and stability in data analysis.

\section{Conclusion}
In this study, we proposed a new measure of asymmetry $\Phi$ based on cosine similarity for square contingency tables dealing with nominal categorical data. 
Under the constraints of $p^c_{ij}$ and $p^c_{ji}$, our proposal was visually demonstrated in the two figures, showing that it provided the simplest and most natural distance-based quantification.
Our proposed measure $\Phi$ offers several key advantages. 
First, the asymmetry of each cell can be represented solely by the odds ${p_{ij}/p_{ji}}$ or ${p_{ji}/p_{ij}}$, making it easily interpretable since it reflects their average. 
Moreover, it exhibits a high sensitivity to asymmetry, enabling the detection of minor departures while also mitigating the influence of extreme asymmetry.
Additionally, the quantification method we adopted corresponds to the Fisher-Rao distance. 
As a result, the measurement by $\Phi$ also functions as a most natural geodesic distance for assessing the difference between probability distributions statistically.
The measure of asymmetry allows for analysis of each square contingency table, separated by confounding factors and other variables, enabling stable and straightforward comparisons between confounding factors that could not be achieved through statistical tests.
Therefore, using our proposed measure to capture transitions is expected to have applications in various fields, including the biological sciences.

\section*{Acknowledgement}
This work was supported by JSPS Grant-in-Aid for Scientific Research (C) Num- ber JP20K03756.

\appendix
\section{Proof of measure's properties}
In this section, we provide a proof for $0 \leq \Phi \leq 1$.
Since $p^c_{ij} = p_{ij}/(p_{ij} + p_{ji})$, it follows that $0 \leq p^c_{ij} \leq 1$.
Thus, considering $p_{ij}^c (1-p_{ij}^c)$, we observe that it takes the maximum value of $1/4$ when $p^c_{ij} = 1/2$ and the minimum value of $0$ when $p_{ij}^c = 0$ or $p_{ij}^c = 1$.
Specifically, since
\begin{equation*}
0 \leq p_{ij}^c (1-p_{ij}^c) \leq \frac{1}{4},
\end{equation*}
we can deduce that
\begin{equation*}
\frac{1}{2} \leq \frac{1}{2}\left\{1 + 2\sqrt{p^c_{ij} (1-p^c_{ij})} \right\} = \left\{ \frac{1}{\sqrt{2}} \left( \sqrt{p^c_{ij}} + \sqrt{p^c_{ji}} \right) \right\}^2  \leq 1.
\end{equation*}
From these inequalities,
\begin{equation*}
\frac{1}{\sqrt{2}} \leq \frac{1}{\sqrt{2}} \left( \sqrt{p^c_{ij}} + \sqrt{p^c_{ji}} \right) \leq 1.
\end{equation*}
Consequently, we obtain
\begin{equation*}
0 \leq \cos^{-1} \left\{ \frac{1}{\sqrt{2}} \left( \sqrt{p^c_{ij}} + \sqrt{p^c_{ji}} \right) \right\} = \cos^{-1} \left( \frac{\sqrt{p_{ij}} + \sqrt{p_{ji}}}{\sqrt{2(p_{ij}+p_{ji})}} \right) \leq \frac{\pi}{4}.
\end{equation*}
The weight $w_{ij}(\bm{p})$ satisfies
\begin{equation*}
w_{ij}(\bm{p}) > 0, \quad \mathop{\sum \sum}_{i \neq j} w_{ij} (\bm{p} ) = 1 .
\end{equation*}
Therefore, the following inequality holds:
\begin{equation*}
0 \leq \frac{4}{\pi} \mathop{\sum \sum}_{i \neq j} w_{ij}(\bm{p}) \cos^{-1} \left( \frac{\sqrt{p_{ij}} + \sqrt{p_{ji}}}{\sqrt{2(p_{ij}+p_{ji})}} \right) \leq 1. 
\end{equation*}

Thus, it can be shown that $\Phi = 0$ when $p^c_{ij} = 1/2 $, i.e., when $p_{ij} = p_{ji}$ for all $i\neq j$.
Additionally, $\Phi = 1$ when $p_{ij}^c = 0$ or $p_{ij}^c = 1$, i.e., when $p_{ij} = 0 \; (p_{ji} >0) $ or $p_{ji} = 0 \;  (p_{ij} >0)$ for all $i \neq j$.

\section{Proof of asymptotic variance for proposed measure}
Let denote $n_{ij}$ as the observed frequency at the intersection of the $i$th row and $j$th column within the table.
Assuming a multinomial distribution applies to the $R \times R$ table, an approximate standard error and a large-sample confidence interval for the measure $\Phi$ are demonstrated using the delta method detailed by \cite{agresti2012categorical}.
The sample version of $\Phi$, denoted as $\hat{\Phi}$, is derived from $\Phi$ with $\{p_{ij} \}$ replaced by $\{ \hat{p}_{ij} \}$, where $\hat{p}_{ij}=n_{ij}/n$ with $n=\sum_{i\neq j} n_{ij}$.
According to the delta method, $ \sqrt{n}(\widehat{\Phi} - \Phi ) $ approaches a normal distribution with a mean 0 and a variance $\sigma^2[\widehat{\Phi}]$ asymptotically.
The asymptotic variance of $\widehat{\Phi}$ is given as follows:
\begin{equation*}
\sigma^2[\widehat{\Phi}] = \mathop{\sum \sum}_{s \neq t} \left( \frac{\partial \Phi}{\partial p_{st}} \right)^2 p_{st} - \left( \mathop{\sum \sum}_{s \neq t}\frac{\partial \Phi}{\partial p_{st}} p_{st}  \right)^2 , 
\end{equation*}
where
\begin{equation*}
\begin{split}
\frac{\partial \Phi}{\partial p_{st}} &= \frac{4}{\pi} \Bigg[\mathop{\sum \sum}_{i \neq j} \left\{ \frac{\partial w_{ij}(\bm{p})}{\partial p_{st}} \cos^{-1} \left( \frac{\sqrt{p_{ij}} + \sqrt{p_{ji}}}{\sqrt{2(p_{ij}+p_{ji})}} \right) \right\} \\
&\quad\quad + w_{st}(\bm{p})\frac{\sqrt{p_{ts}}}{2\sqrt{p_{st}}(p_{st} + p_{ts})} \text{sign}\left(\sqrt{p_{ts}} - \sqrt{p_{st}}\right) \\
&\quad\quad\quad\quad + w_{ts}(\bm{p}) \frac{\sqrt{p_{ts}}}{2\sqrt{p_{st}}(p_{st} + p_{ts})} \text{sign}\left(\sqrt{p_{ts}} - \sqrt{p_{st}}\right)  \Bigg], \end{split}
\end{equation*}
and $\text{sign}(x)$ is a sign function defined as follows:
\begin{equation*}
\text{sign}(x) = \left\{
\begin{array}{rl}
1 & (x > 0),\\
0 & (x = 0), \\
-1 & ( x < 0).
\end{array}
\right.
\end{equation*}
Then, $\widehat{\sigma}[\widehat{\Phi}]/\sqrt{n}$ is an estimated standard error for $\widehat{\Phi}$, where $\widehat{\sigma}^{2}[\widehat{\Phi}]$ is the plug-in estimator of $\sigma^{2}[\widehat{\Phi}]$, that is, $\widehat{\sigma}^{2}[\widehat{\Phi}]$ is obtained by $\sigma^{2}[\widehat{\Phi}]$ with $\{p_{ij}\}$ replaced by $\{\widehat{p}_{ij}\}$.
It should be noted that $ (\sqrt{p_{ts}} - \sqrt{p_{st}})/\vert \sqrt{p_{ts}} - \sqrt{p_{st}} \vert$ is replaced by $\text{sign}(\sqrt{p_{ts}} - \sqrt{p_{st}})$ because the asymptotic confidence interval cannot be computed when there exists at least one pair $(i,j)$ for which $p_{ij} = p_{ji}$ $(i\neq j)$.
Therefore, an approximate $100 ( 1- \alpha)\%$ confidence interval for $\Phi$ is given $\widehat{\Phi} \pm z_{\alpha /2 } \widehat{\sigma}[\widehat{\Phi}]/\sqrt{n}$, where $z_{\alpha}$ in the upper $\alpha \%$ point from the standard normal distribution. 

\section{Real data analysis}
Consider the data by \cite{mukesh2016cambridge}. 
This report is from patients recruited into the trial, which began in April 2003 and ended in June 2007.
Table \ref{table1} and Table \ref{table2} show patient and clinical or photographic assessments of specific normal tissue effects at 2 and 5 years in the Cambridge Breast Intensity-modulated Radiotherapy trial. 
Breast radiotherapy-associated toxicity is often reported using clinical and photographic assessments, and the addition of patient-reported outcome measures (PROMs) is becoming more common.
\cite{mukesh2016cambridge} aims to explore the factors that cause inconsistency between patients' and clinicians' reports.
The article discusses the results of the analysis using the weighted kappa coefficient and Bowker's test, but we also show that the proposed measure can also provide a consideration of symmetric categories.
In addition, as weights to be given to the proposed measure $\Phi$, various weights can be given, but we use (a) $w_{ij} (\bm{p})=1/\{R(R-1)\}$ and (b) $w_{ij} (\bm{p})=(p_{ij}+p_{ji})/(2\sum_{i\neq j} p_{ij})$ in the analysis.

\subsection{Photographic assessment versus Patient-reported outcome}
Let's consider Table \ref{table1}.
The tables show the patient and photographic assessments of breast shrinkage at 2 and 5 years. 
In the report by \cite{mukesh2016cambridge}, the weighted kappa coefficient for Table \ref{table1}(A) was $0.21$, with $P$-value $<0.0001$ in Bowker's test. 
For Table \ref{table1}(B), it was $0.16$ with $P$-value $= 0.0046$.
These results indicate that the agreement between patient and photographic assessment categories is poor, and the symmetric categories that do not match are also not balanced. 
To quantitatively investigate the degree of asymmetry, uneven trends of two different assessments, and whether the asymmetry is resolved between 2 and 5 years, we analyzed the data using our proposed measure.

Table \ref{result1} presents the analysis results, showing the values of estimated measure $\widehat{\Phi}$, the standard errors (SE) of $\Phi$, and $95\%$ confidence intervals (CI) of $\Phi$. 
For each weight, examining the measure's values and confidence intervals for Table \ref{table1}(A) and \ref{table1}(B) indicates, similar to Bowker's test results, no symmetry structure.
Furthermore, while Table \ref{table1}(A) shows a certain degree of departure from symmetry, Table \ref{table1}(B) exhibits a smaller degree of departure. 
Regardless of the weights, the measure's values for Table \ref{table1}(A) are approximately twice that of Table \ref{table1}(B) as seen in Table \ref{result1}. 
However, when checking the confidence intervals for each table, it is observed that the confidence intervals for the two tables overlap regardless of the weights.
From these observations, our proposal, similar to the results of Bowker's test, indicates no symmetry structure in Table \ref{table1}. 
While the heterogeneity in the evaluations appears to be resolved to some extent by the 5 year compared to the 2 year, it cannot be stated with sufficient certainty.

\begin{table}[htp]
\centering
\caption{Patient and photographic assessments of breast shrinkage at 2 and 5 years 
}
\begin{tabular}{crrrr }
\multicolumn{5}{l}{(A) Breast shrinkage/smaller breast - 2 years}  \\
& & \multicolumn{3}{c}{Patient's assessment} \\
\multicolumn{2}{c}{Photographic assessment}  & None & A little & Quite a bit  \\
& None & $288$ & $147$ & $27$  \\ 
& A little & $90$ & $95$ & $33$  \\ 
& Quite a bit & $13$ & $17$ & $14$ \\ \\
\end{tabular}

\begin{tabular}{crrrr}
\multicolumn{5}{l}{(B) Breast shrinkage/smaller breast - 5 years} \\
 & & \multicolumn{3}{c}{Patient's assessment}  \\
\multicolumn{2}{c}{Photographic assessment}  & None & A little & Quite a bit  \\
& None & $181$ & $92$ & $32$ \\ 
& A little & $79$ & $76$ & $24$ \\ 
& Quite a bit & $13$ & $24$ & $13$ \\ 
\end{tabular}
\label{table1}
\end{table}

\begin{table}[htp]
\centering
\caption{Comparison of Table \ref{table1} based on $\Phi$ with two different weights}
{\begin{tabular}{cccc c ccc}
& \multicolumn{3}{c}{Table \ref{table1}(A) } & & \multicolumn{3}{c}{Table \ref{table1}(B)} \\
$w_{ij} (\bm{p})$ & $\widehat{\Phi}$ & SE & $95\%$CI & & $\widehat{\Phi}$ & SE & $95\%$CI  \\
(a) & $0.197$ & $0.047$ & $[0.104, 0.289]$ & & $0.109$ & $0.036$ & $[0.039, 0.178]$ \\ 
(b) & $0.172$ & $0.035$ & $[0.103, 0.241]$ & & $0.079$ & $0.036$ & $[0.008, 0.149]$ \\ 
\end{tabular}}
\label{result1}
\end{table}

\subsection{Clinical assessment versus Patient-reported outcome}
Next, let's consider Table \ref{table2}. 
This table shows the patient and clinical assessments for breast induration at 2 and 5 years. 
In the report by \cite{mukesh2016cambridge}, the weighted kappa coefficient for Table \ref{table2}(A) was $0.05$, with $P$-value $<0.0001$ in Bowker's test. 
For Table \ref{table2}(B), it was $0.10$ with $P$-value $<0.0001$.
These results also indicate that the agreement between patient and clinical assessment categories is poor, and the mismatched categories are not balanced. 
Therefore, we also conducted further analysis using our proposed measure for additional investigation.

Table \ref{result2} presents the analysis results.
Examining the results, it is clear that there are no symmetry structures, similar to the results of Bowker's test.
Additionally, Table \ref{table2}(A) shows a significant departure from symmetry, and Table \ref{table2}(B) also exhibits a substantial departure from symmetry. 
Despite both having a high degree of asymmetry, it can be observed that the measure's value for Table \ref{table2}(A) is approximately twice that of Table \ref{table2}(B) regardless of the weights.
The important point to note is the coverage relationship of the confidence intervals for tables.
For the evaluation of breast induration, the confidence intervals do not overlap regardless of the weights, indicating that the heterogeneity in evaluations is resolved with sufficient accuracy at the 5 year compared to the 2 year.
Another intriguing aspect of Table \ref{result2} is that the measure's values for the two weights are almost unchanged. 
This fact suggests that the asymmetric structure represented directly by $\{p_{ij}/p_{ji}\}$ and $\{p_{ji}/p_{ij}\}$ within the measure is constant, allowing us to assume that the CS model underlies each table.
Referring to Fig.3 of the main paper, it can be inferred that Table \ref{table2}(A) corresponds to approximately $\Delta=0.25$ $($or $ \Delta^{-1}=4)$ for the CS model, and Table \ref{table2}(B) corresponds to approximately $\Delta=0.5$ $($or $\Delta^{-1}=2)$.

From these findings, our proposal suggests the following regarding Table 3: Similar to the results of Bowker's test, there are no symmetry structures, and the degree of departure from symmetry is sufficiently large in both tables. 
Additionally, it can be determined with sufficient certainty that the heterogeneity in evaluations is resolved by the 5 year compared to the 2 year. 
Furthermore, it can be assumed that the CS model exhibits in the tables.
In the evaluation of breast induration at the 2 year, it can be said that clinicians tend to evaluate the symptoms as approximately $\Delta=0.25$ $($or $\Delta^{-1}=4)$ times more severe than patients.
On the other hand, at the 5 year, there is a tendency for patients to misinterpret symptoms that clinicians evaluated as ``A little'', approximately $\Delta=0.5$ $($or $\Delta^{-1}=2)$ times more frequently.

\begin{table}[htp]
\centering
\caption{Patient and clinical assessments of breast induration at 2 and 5 years 
}
\begin{tabular}{crrrr}
\multicolumn{5}{l}{(A) Breast induration/hardness - 2 years}\\
& & \multicolumn{3}{c}{Patient's assessment} \\
\multicolumn{2}{c}{Clinical assessment}  & None & A little & Quite a bit  \\
& None & $85$ & $53$ & $13$  \\ 
& A little & $187$ & $128$ & $33$  \\ 
& Quite a bit & $128$ & $108$ & $43$ \\ \\
\end{tabular}

\begin{tabular}{crrrr}
\multicolumn{5}{l}{(B) Breast induration/hardness - 5 years} \\
& & \multicolumn{3}{c}{Patient's assessment}  \\
\multicolumn{2}{c}{Clinical assessment}  & None & A little & Quite a bit  \\
& None & $121$ & $37$ & $70$ \\ 
& A little & $157$ & $81$ & $52$ \\ 
& Quite a bit & $70$ & $14$ & $19$ \\ 
\end{tabular}

\label{table2}
\end{table}

\begin{table}[htp]
\centering
\caption{Comparison of Table \ref{table2} based on $\Phi$ with two different weights}
{\begin{tabular}{cccc c ccc}
& \multicolumn{3}{c}{Table \ref{table2}(A) } & & \multicolumn{3}{c}{Table \ref{table2}(B)} \\
$w_{ij} (\bm{p})$ & $\widehat{\Phi}$ & SE & $95\%$CI & & $\widehat{\Phi}$ & SE & $95\%$CI  \\
(a) & $0.447$ & $0.029$ & $[0.391, 0.503]$ & & $0.272$ & $0.030$ & $[0.212, 0.331]$ \\ 
(b) & $0.434$ & $0.028$ & $[0.378, 0.489]$ & & $0.270$ & $0.028$ & $[0.216, 0.324]$ \\ 
\end{tabular}}
\label{result2}
\end{table}


\newpage

\bibliographystyle{apacite}
\bibliography{paper-ref}

\end{document}